\documentclass[a4paper]{amsart}

\usepackage{graphics,amssymb,enumerate}
\usepackage{hyperref}
\usepackage[dvips]{graphicx}
\usepackage[all]{xy}

\def\bd{\begin{displaymath}}
\def\ed{\end{displaymath}}
\def\be{\begin{equation}}
\def\ee{\end{equation}}
\def\p{\partial}

\newtheorem{theorem}{Theorem}

\begin{document}

\title[Realization Volterra]{A symplectic realization of the Volterra lattice}

\author{M. A.  Agrotis}
\address{Department of Mathematics and Statistics\\
University of Cyprus\\
P.O.~Box 20537, 1678 Nicosia\\Cyprus} \email{damianou@ucy.ac.cy}

\author{P. A.~Damianou}
\address{Department of Mathematics and Statistics\\
University of Cyprus\\
P.O.~Box 20537, 1678 Nicosia\\Cyprus} \email{damianou@ucy.ac.cy}

\author{ G. Marmo}
\address{Dipartimento di Scienze Fisiche, Universita' Federico II di
Napoli, Italy INFN, Sezione di Napoli } \email{damianou@ucy.ac.cy}

\thanks{One of the authors M.A.A would like to thank
the Cyprus Research Promotion Foundation for support through the
grant CRPF0504/03. }

\begin{abstract}
We examine the multiple Hamiltonian structure  and construct a
symplectic realization of the Volterra model. We rediscover the
hierarchy of invariants, Poisson brackets and master symmetries
via the use of a recursion operator. The rational Volterra bracket
is obtained using a negative recursion operator.
\end{abstract}

\date{}
\maketitle

\section{Introduction} \label{intro}

The Volterra model, also known as KM system is defined by

\begin{equation}
\dot u_i = u_i(u_{i+1}-u_{i-1}) \qquad i=1,2, \dots,n, \label{a1}
\end{equation}
where $u_0 \!= u_{n+1} \!=0$. It was studied originally by
Volterra in \cite{volterra} to describe population evolution in a
hierarchical system of competing species. It was first solved by
Kac and van-Moerbeke in \cite{kac}, using a discrete version of
inverse scattering due to Flaschka \cite{flaschka}. In
\cite{moser2} Moser gave a solution of the system using the method
of continued fractions and in the process he constructed
action-angle coordinates. Equations (\ref{a1}) can be considered
as a finite-dimensional approximation of the Korteweg-de Vries
(KdV) equation. They also appears in the discretization of
conformal field theory; the Poisson bracket  for this system can
be thought as a lattice generalization of the Virasoro algebra
\cite{fadeev2}. The variables $u_i$  are an intermediate step in
the construction of the action-angle variables for the Liouville
model on the lattice. This system has a close connection with the
Toda lattice,
\begin{eqnarray*}
\dot{a}_i=a_i (b_{i+1}-b_{i}) \quad& i=1,\ldots,n-1 \\
\dot{b}_i=2(a_i^2-a_{i-1}^2 & i=1,\ldots,n.
\end{eqnarray*}
In fact, a transformation of H\'{e}non connects the two systems:
\begin{eqnarray*}
a_i=-\frac{1}{2}\sqrt{u_{2i}u_{2i-1}} & i=1,\ldots,n-1 \\
b_i=\frac{1}{2}(u_{2i-1}+u_{2i-2}) \quad& i=1,\ldots,n.
\end{eqnarray*}
We note that the number of variables for the Toda lattice is odd
and therefore we restrict our attention to the Volterra  system
with an odd number of variables. The Volterra system is usually
associated with a simple Lie algebra of type $A_n$. Bogoyavlensky
generalized this system for each simple Lie algebra and showed
that the corresponding systems are also integrable. See
\cite{bog1,bog2} for more details. The relation between Volterra
and Toda systems is also examined in \cite{damianou3}.

The Hamiltonian description of system (\ref{a1}) can be found in
\cite{fadeev} and \cite{damianou2}. We will follow
\cite{damianou2} and use the Lax pair of that reference. The Lax
pair is given by

\begin{equation*}
\dot{L}=[B, L],
\end{equation*}
where

\begin{equation*}
 L= \begin{pmatrix} u_1 & 0 & \sqrt{u_1 u_2} & 0 & \dots & \  & 0 \cr 0
& u_1 +u_2 & 0& \sqrt{u_2 u_3}  & & & \vdots \cr
 \sqrt{u_1 u_2} & 0 & u_2 +u_3 & &  \ddots & &  \cr
 0 & \sqrt{u_2 u_3} & &  & & & \cr
  \vdots & & \dots & & & & \sqrt{u_{n-1} u_n} \cr
 & & & & & u_{n-1}+u_n & 0 \cr
 & & & & \sqrt{u_{n-1} u_n} &0& u_n \end{pmatrix}
\end{equation*}
and

\begin{equation*}
 B=\begin{pmatrix} 0 & 0 & { 1 \over 2} \sqrt{u_1 u_2} & 0 & \dots & \
& 0 \cr
 0 & 0 & 0&{ 1 \over 2} \sqrt{u_2 u_3}  & & & \vdots \cr
 -{ 1\over 2} \sqrt{u_1 u_2} & 0 & 0 & &  \ddots & &  \cr
 0 & -{ 1\over 2} \sqrt{u_2 u_3} & &  & & & \cr
  \vdots & & \dots & & & & {1 \over 2} \sqrt{u_{n-1} u_n} \cr
 & & & & & 0 & 0 \cr
 & & & &-{ 1 \over 2}  \sqrt{u_{n-1} u_n} &0& 0 \end{pmatrix}.
 \end{equation*}
This is an example of an isospectral deformation; the entries of
$L$ vary over time but the eigenvalues  remain constant. It
follows that the  functions $ H_i={1 \over i} {\rm Tr} \, L^i$ are
constants of motion.  We note that \bd H_1= 2 \sum_{i=1}^n u_i \
\ed corresponds to the total momentum and \bd H_2= \sum_{i=1}^n
u_i^2+ 2 \sum_{i=1}^{n-1} u_i u_{i+1}\ed is the Hamiltonian.

Following \cite{damianou2} we define the following quadratic
Poisson bracket, \bd \{u_i, u_{i+1} \}=u_i u_{i+1}, \ed and all
other brackets equal to zero. We denote this bracket by $\pi_2$.
For this bracket det$L$ is a Casimir and the eigenvalues of $L$
are in involution. Of course, the functions $H_i$ are also in
involution. Taking the function $\sum_i^{n} u_i $ as the
Hamiltonian we obtain equations (\ref{a1}). This bracket can be
realized from the second Poisson bracket of the Toda lattice by
setting the momentum variables equal to zero \cite{fadeev}.

In \cite{damianou2} one also finds a cubic Poisson bracket which
corresponds to the second KdV bracket in the continuum limit. It
is defined by the formulas,

\begin{equation*}
\begin{array}{rcl}
\{ u_i, u_{i+1} \}&=& u_i u_{i+1} (u_i+ u_{i+1}) \\
\{ u_i, u_{i+2} \} &=& u_i u_{i+1} u_{i+2} \, ,
\end{array}
\end{equation*} all other brackets are zero. We denote this bracket by
$\pi_3$.  In this bracket we still have involution of invariants.
We also have Lenard type relations of the form \bd \pi_3\, \nabla
H_i = \pi_2 \, \nabla H_{i+1} . \ed

In \cite{damianou2} appears a  bracket that is homogeneous of
degree one, a rational bracket constructed using a master
symmetry. This bracket, denoted by $\pi_1$, has ${\rm Tr} L$ as
Casimir and the Hamiltonian is $H_2={ 1\over 2} {\rm Tr} L^2$. The
definition of the bracket is the following. We define the master
symmetry $Y_{-1}$ to be

\bd Y_{-1}=\sum_{i=1}^n  f_i { \p \over \p u_i}  \ , \ed where the
$f_i$ are determined recursively as follows,

\bd f_1=-1, \quad f_{2i}=\frac{u_{2i}} {u_{2i-1}} f_{2i-1} , \quad
f_{2i-1}=-f_{2i-2} -1  . \ed Taking the Lie derivative of $\pi_2$
in the direction of $Y_{-1}$ we obtain $\pi_1$, a Poisson bracket
that is homogeneous of degree 1. For $n=5$, $\pi_1$ takes the
form:

\begin{equation}
\begin{array}{lll}
 \{u_1, u_2\} = u_2 \qquad \{u_1, u_3\} =-u_2  \quad \qquad \{u_1
, u_4\}= { u_2 u_4 \over u_3 } \qquad \{u_1 , u_5 \} =-{ u_2 u_4
\over u_3 } \\
  \{u_2, u_3 \}=u_2 \qquad \{u_2, u_4\} = -{ u_2 u_4 \over u_3
}  \hskip .07cm \qquad \{ u_2, u_5 \}={u_2 u_4 \over u_3 } \label{bracketpi1} \\
 \{u_3, u_4\} = u_4 \qquad \{u_3, u_5 \}=-u_4   \quad \qquad
\{u_4, u_5 \}=u_4.
\end{array}
\end{equation}

\noindent In this paper we rediscover this bracket using a
recursion operator. The higher Poisson brackets are constructed
using a sequence of master symmetries $Y_i, \: \: i=-1,0,1,
\dots$. We define $Y_0$ to be the Euler vector field

\bd Y_0=\sum_{i=1}^n u_i {\partial \over \partial u_i} \ . \ed The
explicit formula for $Y_1$ is \bd Y_1=\sum_{i=1}^n U_i {\p \over
\p u_i} \ , \ed where \bd U_i=(i+1) u_i u_{i+1} +u_i^2+(2-i)
u_{i-1} u_i \ . \ed

\noindent It is easily checked that the bracket $\pi_2$ is
obtained from $\pi_1$ by taking the Lie derivative in the
direction of $Y_1$. Similarly, the Lie derivative of $\pi_2$ in
the direction of $Y_1$ gives $\pi_3$.

The brackets $\pi_1$, $\pi_2$ and $\pi_3$ are just the beginning
of an infinite hierarchy constructed in \cite{damianou2} using
master symmetries. We quote the result:

\begin{theorem} \label{th1}
There exists a sequence of Poisson tensors $\pi_j$ and a sequence
of master symmetries $Y_j$ such that:

\smallskip

\noindent {\it i)} $\pi_j$ are all Poisson.

\smallskip

\noindent {\it ii)} The functions $H_i$  are in involution
 with respect to all of the $\pi_j$.

\smallskip

 \noindent
 {\it iii)} $Y_i (H_j) =(i+j) H_{i+j} $.

\smallskip

 \noindent
{\it iv)} $L_{Y_i} \pi_j =(j-i-2) \pi_{i+j} $.

\smallskip

\noindent {\it v)} $[Y_i, \ Y_j]=(j-i)Y_{i+j}$.

\smallskip

\noindent {\it vi)} $\pi_j \nabla H_i =\pi_{j-1} \nabla H_{i+1} $,
where $\pi_j$ denotes  the Poisson matrix  of the tensor $\pi_j$.
\end{theorem}

In this paper we prove the results of Theorem 1 using a different
approach. Namely, we construct a recursion operator in a
symplectic space, define all master symmetries, invariants and
Poisson brackets using results of Magri and Oevel and then project
to the space of $u$ variables.

\section{Master Symmetries and a Theorem of Oevel}

We recall the definition and basic properties of master symmetries
following Fuchssteiner \cite{fuchssteiner}. Consider a
differential equation on a manifold $M$  defined by a vector field
$\chi$.
 We are mostly interested in the case where $\chi$ is a Hamiltonian vector
field. A
 vector field $Z$ is a   symmetry of the equation  if
\begin{equation*}
[Z, \chi]=0  .
\end{equation*}
A vector field $Z$ is called a master symmetry if
\begin{equation*}
[[Z, \chi], \chi]=0 ,
\end{equation*}
but
\begin{equation*}
[Z, \chi] \not= 0  .
\end{equation*}

\noindent Master  symmetries were first introduced by Fokas and
Fuchssteiner in \cite{fokas1}  in connection with the Benjamin-Ono
Equation.

A bi-Hamiltonian system is defined by specifying two Hamiltonian
functions $H_1$, $H_2$ and two Poisson tensors $\pi_1$ and
$\pi_2$, that give rise to the same Hamiltonian equations. Namely,
$
\pi_1 \nabla H_2=\pi_2 \nabla H_1.
$ The notion of bi-Hamiltonian structures is due to Magri
\cite{magri}. Suppose that we have a bi-Hamiltonian system defined
by the Poisson tensors $\pi_1$, $\pi_2$ and the Hamiltonians
$H_1$, $H_2$.
 Assume that $\pi_1$ is symplectic.  We define
the recursion operator ${\mathcal R} = \pi_2 \pi_1^{-1}$,  the
higher flows
\begin{equation*}
\chi_{i} = {\mathcal R}^{i-1} \chi_1 \ ,
\end{equation*}
and the higher order Poisson tensors \bd \pi_i = {\mathcal
R}^{i-1} \pi_1 \ . \ed

\noindent For a non-degenerate bi-Hamiltonian system, master
symmetries can be generated using a method due to Oevel
\cite{oevel2}.

\begin{theorem} \label{th2}
Suppose that   $X_0$ is a conformal symmetry for both  $\pi_1$,
$\pi_2$ and $H_1$, i.e.  for some scalars $\lambda$, $\mu$, and
$\nu$ we have \bd {\mathcal L}_{X_0} \pi_1= \lambda \pi_1,
\quad{\mathcal L}_{X_0} \pi_2 = \mu \pi_2, \quad {\mathcal
L}_{X_0} H_1 = \nu H_1  . \ed Then the vector fields $X_i =
{\mathcal R}^i X_0$ are master symmetries and we have,
\begin{eqnarray*}
 (a) \ {\mathcal L}_{X_i} H_j = (\nu +(j-1+i) (\mu -\lambda))
H_{i+j}
\\
 (b) \ {\mathcal L}_{X_i} \pi_j = (\mu +(j-i-2) (\mu -\lambda))
\pi_{i+j} \\
 (c) \ [X_i, X_j]= (\mu - \lambda) (j-i) X_{i+j}  \ .
\end{eqnarray*}
\end{theorem}

\section{Symplectic realization }

We define the following transformation from ${\bf R}^{2n}$ to
${\bf R}^{2n-1}$,
\begin{eqnarray}
u_{2i-1} =  - e^{p_i} & i=1,\ldots,n  \ ,\nonumber \\
u_{2i} =  e^{q_{i+1}- q_i}  \qquad& i=1,\ldots,n-1
\label{u-variables} \ .
\end{eqnarray}

\noindent The Hamiltonian in $(q,p)$ coordinates is given by

\be h_1= - \sum_{i=1}^n e^{p_i} +\sum_{i=1}^{n-1}
e^{q_{i+1}-q_i}. \label{b2} \ee

\noindent It is straightforward to check that Hamilton's equations
for (\ref{b2}) correspond in the $u-$space to the KM-system
(\ref{a1}) via the mapping (\ref{u-variables}). The symplectic
bracket in $(q,p)$ coordinates corresponds to the quadratic
bracket $\pi_2$. For this reason we will denote the standard
symplectic bracket in ${\bf R}^{2n}$ by $J_2$.  Our purpose is to
define a bracket $J_3$ in ${\bf R}^{2n}$ which is mapped to
$\pi_3$ under the transformation (\ref{u-variables}). The idea of
the construction is to lift the master symmetry $Y_1$ from the
$u-$space up to the $(q,p)-$space and obtain a vector field which
we denote by $X_1$. The new bracket $J_3$ will be defined as the
Lie derivative of $J_2$ in the direction of $X_1$. One possible
definition for $X_1$ is the following: \bd X_1 = \sum_{i=1}^n A_i
{\p \over \p q_i}  + \sum_{i=1}^n B_i{ \p \over \p p_i} \, \ed
where,
\begin{eqnarray*}
A_i = -e^{p_1} -\sum_{j=2}^{i-1} e^{p_j} +(1-2 \, i) e^{p_i}
+\sum_{j=1}^{i-1}
e^{q_{j+1} -q_j}  \qquad& i=1,2, \dots, n \ ,\\
B_i = 2 \, i \, e^{q_{i+1} -q_i} -e^{p_i} +(3-2 \, i) e^{q_i
-q_{i-1} } & i=1,2, \dots, n \label{b1} \ .
\end{eqnarray*}
We note that in the summations if an index is not defined then we
ignore that whole term.

Taking the Lie derivative of the symplectic bracket $J_2$ in the
direction of $X_1$ we obtain the Poisson bracket $J_3$,
\begin{eqnarray}
\{ q_i, q_{j} \}  =  e^{p_j} &   1 \le j \le i-1 \le
n-1  \label{J3bracket} \\
 \{ q_i, p_i \} =  -e^{p_i} + e^{q_i -q_{i-1} } & i= 1,
\dots, n \nonumber \\
\{ q_{i}, p_j \} =  e^{q_j - q_{j-1}}- e^{q_{j+1}- q_{j}} \qquad&
1  \le j \le i-1 \nonumber \\
\{ p_i , p_{i+1} \} =  e^{q_{i+1} -q_{i}} & i=1, \dots, n-1
\nonumber
\end{eqnarray}
The Jacobi identity is straightforward to check. There are four
cases (three $p$, three $q$, two $p$ one $q$, and two $q$ one
$p$). Two of the cases are trivial and the other two can be
broken-up to at most five subcases.

Let ${J}_2$ be the symplectic bracket with Poisson matrix

\begin{equation*}
{J}_2 = \begin{pmatrix} 0 &  I \cr
                      -I &   0 \end{pmatrix},
\end{equation*} where $I$ is the $n \times n$ identity matrix.
The bracket $J_2$ is mapped precisely to the bracket $\pi_2$ under
transformation (\ref{u-variables}), and $J_3$ corresponds to
$\pi_3$. We define a recursion operator as follows:

\begin{equation*}
{\mathcal R}=J_3 J_2^{-1} .
\end{equation*}
This operator raises degrees and we therefore   call it the
positive Volterra operator. In $(q,p)$ coordinates, the  symbol
$\chi_i$ is a shorthand for  $\chi_{h_i}$. It is generated, as
usual, by

\begin{equation*} \chi_i = {\mathcal R}^{i-1} \chi_1 . \end{equation*}
For example,

\begin{equation*}
h_2=\frac{1}{2} \sum_{i=1}^{n}
e^{2p_i}+\frac{1}{2}\sum_{i=1}^{n-1}
e^{2(q_{i+1}-q_{i})}-\sum_{i=1}^{n-1}
(e^{p_{i}}+e^{p_{i+1}})e^{q_{i+1}-q_i}.
\end{equation*}
Note that $h_2$ corresponds under mapping (\ref{u-variables}) to a
constant multiple of $H_2=\frac{1}{2} {\rm Tr} \, (L)^2$. In a
similar fashion we obtain the higher order Poisson tensors

\bd J_i = {\mathcal R}^{i-2} J_2 \qquad i=3,4, \dots . \ed

\noindent We finally define the conformal symmetry \bd
X_0=\sum_{i=1}^n i {\partial \over
\partial q_i} +\sum_{i=1}^n {\partial \over
\partial p_i} \ . \ed

\noindent The Poisson tensors $J_2, J_3$ and the functions
$h_1,h_2$ define a bi-Hamiltonian pair. Namely, $J_2 \nabla
h_2=J_3 \nabla h_1$. We note that $J_3$ is automatically
compatible with $J_2$ since it is constructed using a master
symmetry (see \cite{damianou4} p.5518). It is straightforward to
verify that

\begin{equation*} {\mathcal L}_{X_0} J_2=0, \quad
{\mathcal L}_{X_0} J_3= J_3, \quad {\mathcal L}_{X_0}(h_1)=h_1 .
\end{equation*}
Consequently, $X_0$ is a conformal symmetry for $J_2$, $J_3$ and
$h_1$. The constants appearing in Oevel's Theorem are $\lambda=0$,
$\mu=1$ and $\nu=1$. Therefore, we end up with the following
deformation relations:

\bd [X_i, h_j]= (i+j)h_{i+j} \ed

\bd L_{X_i}  J_j = (j-i-2) J_{i+j} \ed

\bd
 [ X_i, X_j ]  = (j-i) X_{i+j}  \ .
\ed

\noindent Projecting to the $u-$space under mapping
(\ref{u-variables}) we obtain relations (iii)-(v) of Theorem
\ref{th1}. Statements (i) and (ii) of Theorem \ref{th1} follow
easily from properties of the recursion operator.

\section{The negative Volterra hierarchy}

In this section we describe how the first bracket $\pi_1$ is
obtained via the use of the negative operator. The negative
operator was introduced in \cite{damianou2a} in connection with
the Toda lattice. We define $J_1$ as follows:

\begin{equation*}
J_1=\mathcal{N}J_2,\;\; \mbox{where} \;\; \mathcal{N}=J_2
J_3^{-1}. \end{equation*} We then project the $J_1$ bracket to the
$u-$space using transformation (\ref{u-variables}) to obtain the
bracket $\pi_1$. We illustrate in detail the case $n=5$.

We consider the Volterra model in ${\bf R}^6$ with coordinates
$(q_1,q_2,q_3,p_1,p_2,p_3)$. Transformation (\ref{u-variables}) is
given by,

\begin{equation}
u_1=-e^{p_1}, \quad u_3=-e^{p_2}, \quad u_5=-e^{p_3}, \quad
u_2=e^{q_2-q_1}, \quad u_4=e^{q_3-q_2}. \label{d1}
\end{equation}

\bd J_2=\begin{pmatrix} 0 & I_3 \cr
                -I_3 & 0\end{pmatrix}, \ed where $I_3$ is the $3 \times 3$ identity matrix,
                 and $J_3$ is the Poisson matrix (\ref{J3bracket}),

\bd  J_3=\begin{pmatrix} 0 & -e^{p_1} & -e^{p_1} & -e^{p_1} & 0 &
0 \cr
                    e^{p_1} & 0 & -e^{p_2} & -e^{q_2-q_1} &
                    -e^{p_2} + e^{q_2-q_1} & 0 \cr
                    e^{p_1} & e^{p_2} & 0 & -e^{q_2-q_1} &
                    e^{q_2-q_1}-e^{q_3-q_2} & -e^{p_3}+e^{q_3-q_2}
                    \cr e^{p_1} & e^{q_2-q_1} & e^{q_2-q_1} & 0 &
                    e^{q_2-q_1} & 0 \cr 0 & e^{p_2}-e^{q_2-q_1} &
                    -e^{q_2-q_1}+e^{q_3-q_2} &  -e^{q_2-q_1} & 0
                    & e^{q_3-q_2} \cr 0 & 0 & e^{p_3}-e^{q_3-q_2} &
                    0 & -e^{q_3-q_2} & 0\end{pmatrix}
                    .
\ed One can find the matrix $J_1$,

\begin{eqnarray*}
 (J_1)_{1,2}=\frac{1}{D} e^{p_1} (e^{p_3}-e^{q_3-q_2}) &
(J_1)_{1,3}=\frac{1}{D} e^{p_1} (e^{p_2}-e^{q_3-q_2}) \\
 (J_1)_{1,4}=\frac{1}{D} e^{q_2-q_1} [(e^{p_3} - e^{q_3-q_2}) -
e^{p_2} e^{p_3}] \qquad& (J_1)_{1,5}= \frac{1}{D} e^{p_1}
e^{q_3-q_2} \\
 (J_1)_{1,6}=-\frac{1}{D} e^{p_1} e^{q_3-q_2} &
(J_1)_{2,3}=\frac{1}{D} e^{p_1} e^{p_2} \\
 (J_1)_{2,4}=\frac{1}{D} e^{q_2-q_1} (e^{p_3} - e^{q_3-q_2}) &
(J_1)_{2,5}=\frac{1}{D} e^{p_1} (e^{q_3-q_2} - e^{p_3}) \\
 (J_1)_{3,4}=-\frac{1}{D} e^{q_2-q_1}e^{q_3-q_2} &
(J_1)_{3,5}=\frac{1}{D} e^{p_1}e^{q_3-q_2} \\
 (J_1)_{3,6}=-\frac{1}{D} e^{p_1} e^{p_2} &
(J_1)_{4,5}=-\frac{1}{D} e^{p_3} e^{q_2-q_1} \\
 (J_1)_{4,6}=\frac{1}{D} e^{q_2-q_1} e^{q_3-q_2}&
(J_1)_{5,6}=-\frac{1}{D} e^{p_1} e^{q_3-q_2},
\end{eqnarray*}
where $D=e^{p_1} e^{p_2} e^{p_3}$. We note that $D$ corresponds in
the $u-$space to the square root of $\det(L)$. The projection of
$J_1$ to the $u-$space under transformation (\ref{d1}) is
precisely the bracket $\pi_1$ given in equation
(\ref{bracketpi1}), e.g.
\begin{eqnarray*}
 \{u_1,u_2\}& =\{-e^{p_1},e^{q_2-q_1}\}=-e^{p_1}e^{q_2-q_1}
(\{p_1,q_2\}-\{p_1,q_1\}) \\
 &=-\frac{e^{p_1}e^{q_2-q_1}}{e^{p_1}e^{p_2}e^{p_3}} \left[
e^{q_2-q_1}(-e^{p_3}+e^{q_3-q_2})+e^{q_2-q_1}
e^{p_3}-e^{q_2-q_1}e^{q_3-q_2}-e^{p_3}e^{p_2} \right] \\
 &= e^{q_2-q_1} = u_2.
\end{eqnarray*}
Using the recursion operator $\mathcal{N}$ we can construct the
negative Volterra hierarchy, i.e. $J_{i-1}=\mathcal{N}J_{i}, \;
i=1,0,-1,-2,\ldots$. Using the same method of proof as in
\cite{damianou2a} one can easily show that the conclusions of
Theorem \ref{th1} hold for any integer value of the index. For
example, for $i=1$ we obtain a Poisson bracket $J_0$, which
projected to the $u-$space gives a rational Poisson bracket of
degree zero, $\pi_0$. In the case of the Volterra model in
${\bf{R}}^4$ one can find that $\pi_0$ is given by,

\begin{eqnarray*}
&& \{ u_1,u_2 \} =  \frac{u_2(u_2+u_3)}{u_1 u_3} \quad \qquad \{
u_3,u_1 \} =
\frac{u_2(u_1+u_2+u_3)}{u_1 u_3} \\
&& \{ u_2,u_3\}=\frac{u_2(u_2+u_1)}{u_1 u_3} \ .
\end{eqnarray*}

\section{Conclusions}
This paper contains three main ingredients. The first, consists of
the odd--dimensional space of the Volterra model together with its
multiple Hamiltonian structures. These results are not new but
they are derived in this paper using an entirely new approach. The
quadratic and cubic brackets $\pi_2$ and $\pi_3$ are contained
implicitly in the book or Fadeev  and Takhtajan \cite{fadeev}. The
rational linear bracket $\pi_1$ and the rest of the hierarchy were
first computed in \cite{damianou2} using master symmetries.

The second part is a realization of the model in  a symplectic
space. We define a Hamiltonian system in $(q,p)$ coordinates, we
compute master symmetries and a second Poisson bracket which
defines a bi--Hamiltonian pair. We then use the recursion operator
to produce the infinite hierarchy.  All the results here are new.

The third part is  the mapping which connects the two spaces and
the two systems. It is a mapping from an even $2n$ dimensional
space to an odd $2n-1$ dimensional space. This symplectic
realization is also new. We have to mention that there is another
symplectic realization of the model which goes back to Volterra.
However,  the map is from a $4n-2$ to a $2n-1$ space, see e.g.
\cite{fernandes}. Due to the big difference in dimension the
results of the present paper will be difficult to duplicate for
that particular realization.

\end{document}